\documentclass[a4paper,11pt]{article}
\usepackage{pos}
\usepackage{tikz}

\newcommand{\sqrts}{\sqrt{s_\textrm{NN}}}
\newcommand{\kompost}{K\o{}MP\o{}ST}

\title{Electromagnetic probes in heavy-ion collisions: progress and open questions}

\author[a,b]{Jean-François Paquet}

\affiliation[a]{Department of Physics and Astronomy, Vanderbilt University,\\Nashville TN 37212}

\affiliation[b]{Department of Mathematics, Vanderbilt University,\\Nashville TN 37212}

\emailAdd{jean-francois.paquet@vanderbilt.edu}

\abstract{I present an overview of photon and dilepton production in heavy-ion collisions, highlighting recent progress and ongoing challenges, with focus on hard initial scatterings, pre-equilibrium electromagnetic emission, as well as thermal and hadronic production. The potential of photons and dileptons to probe low-energy collisions is discussed briefly.}

\FullConference{
 26-31 March 2023 \\
 Aschaffenburg, Germany\\}

\begin{document}
\maketitle

\section{Introduction}

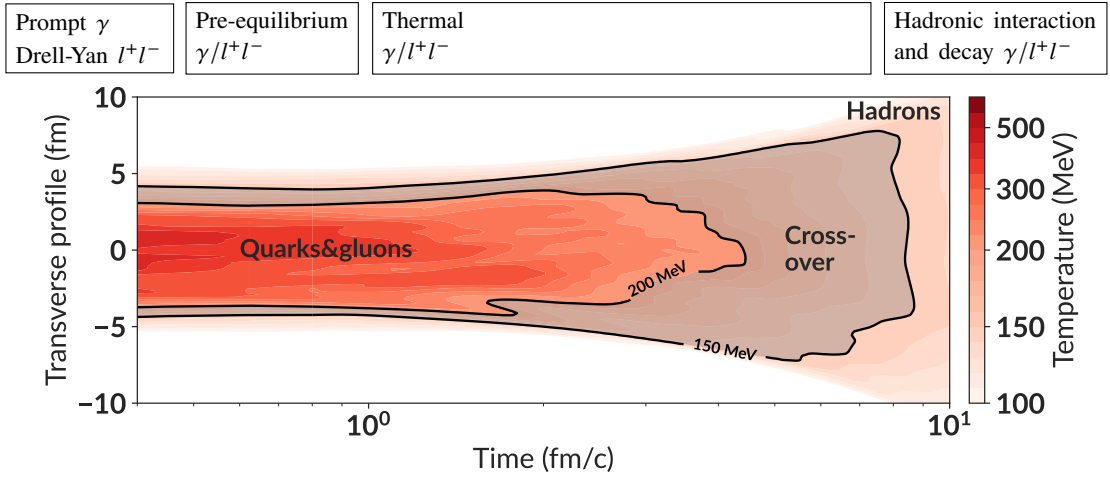
\begin{figure}[hbtp]
	\centering
	\vspace{1.5cm}
	\begin{tikzpicture}
		\node at (0,0) {\includegraphics[width=.9\textwidth]{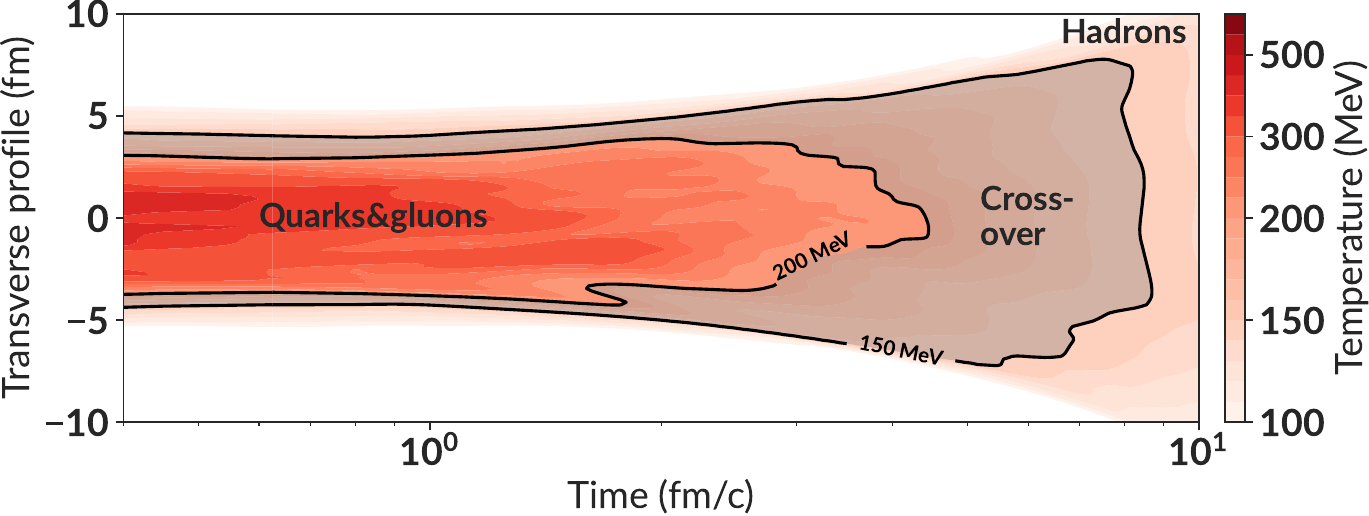}};
		\node[overlay,text width=1.95cm,draw] at (-6.2,3.2) {\footnotesize Prompt $\gamma$ \\[-0.6ex] Drell-Yan $l^+l^-$};
		\node[overlay,text width=2.01cm,draw] at (-3.8,3.2) {\footnotesize  Pre-equilibrium \\[-0.8ex] $\gamma/l^+l^-$};
		\node[overlay,text width=6.3cm,draw] at (0.8,3.2) {\footnotesize  Thermal \\[-0.8ex] $\gamma/l^+l^-$};
		\node[overlay,text width=2.67cm,draw] at (5.72,3.2) {\footnotesize  Hadronic interaction \\[-0.8ex] and  decay $\gamma/l^+l^-$};
	\end{tikzpicture}
	\caption{Temperature profile for a single Au-Au collision at $\sqrts=200$~GeV, as a function of one of the transverse directions and longitudinal proper time $\tau=\sqrt{t^2-z^2}$. Some sources of electromagnetic radiation are indicated at the top, approximately in the order in which they are produced. The temperature profile was obtained by combining IP-Glasma initial conditions, \kompost{} pre-equilibrium evolution and viscous relativistic hydrodynamics as described in Ref.~\cite{Gale:2021emg}.}
		\label{fig:temperature_profile}
\end{figure}

Ultrarelativistic heavy-ion collisions produce a quark-gluon plasma that is electrically charged but whose dynamics is governed by the strong interaction. 
Electromagnetic radiation is emitted at every stage of the collision, as illustrated in Figure~\ref{fig:temperature_profile}. 
Photons and dileptons produced at these different stages escape the medium with negligible interactions, and as such the inclusive electromagnetic signal measured in nuclear collisions is a superposition of all possible production mechanisms (see Refs~\cite{David:2019wpt} and \cite{Geurts:2022xmk} for recent overviews of photon and dilepton production in heavy-ion collisions).

In terms of sheer number of produced photons and dileptons, the overwhelming majority originates from hadronic decays (e.g. $\pi^0 \to \gamma \gamma$).
Nevertheless, different mechanisms of photon and dilepton production tend to contribute to different ranges of momentum and invariant mass; focusing on certain bins of momentum and invariant mass can help increase one's sensitivity to photons and dileptons produced at a certain stage of the collision.
Production channels can also be partly isolated by looking at correlations of photons and dileptons with other particles.
A complementary approach is to subtract or exclude photons or dileptons originating from the overwhelming hadronic decay channels, allowing other sources of electromagnetic radiation to be visible.

Photons and dileptons can provide a stringent test of our understanding of nuclear collisions: they probe the parton distribution of the nuclei, the approach to thermal and chemical equilibrium of the system, the plasma's properties as it expands, its recombination into hadrons, and the final interaction and decays of these hadrons. 
On the other hand, photons and dileptons tend to be challenging to measure, and experimental uncertainties are often considerable compared to measurements of \emph{hadronic} multiplicities, mean transverse momenta, anisotropic coefficients $v_n$, etc.  Increasing the accuracy of electromagnetic measurements will provide valuable insights into heavy-ion collisions.

\section{Evaluating photon and dilepton production}

\label{sec:micro_macro}

\paragraph{Microscopic approach} In certain cases, the processes that lead to electromagnetic radiation can be described microscopically. One example is the production of photons and dileptons in initial hard interactions of the partons from each nucleus, which lead to ``prompt photons'' $\gamma$ and ``Drell-Yan dileptons'' $l^+ l^-$ through processes such as quark-antiquark annihilation
 ($q \bar{q} \to g \gamma$ or $q \bar{q} \to l^+ l^-$). 
Another example is the interaction of individual hadrons in the later stage of the collision, which can be described with transport approaches (SMASH, UrQMD, JAM, \ldots); at each collision, one can compute the probability of producing photons or dileptons (e.g. $\pi + \pi \to \rho + \gamma$; see Section~\ref{sec:hadronic}). Other examples include PHSD, which evaluates photons and dileptons microscopically from quark and gluon degrees of freedom~\cite{Linnyk:2015rco}, as well as electromagnetic radiation from hadronization (see Ref.~\cite{Fujii:2022hxa} and references therein).

\paragraph{Macroscopic approach --- local equilibrium}

In other cases, the degrees of freedom that produce electromagnetic radiation are described macroscopically. For example, hydrodynamic simulations of the deconfined plasma produced in heavy-ion collisions imply that the energy-momentum tensor $T^{\mu\nu}$ encodes all available information about the energy and momentum of the underlying degrees of freedom. 
The information from the energy-momentum tensor can be combined with a macroscopic rate of electromagnetic radiation per volume of plasma, $E d^3 \Gamma/d k^3$.
This macroscopic approach works well if the underlying degrees of freedom are \emph{in} local thermal equilibrium, in which case the energy-momentum tensor can be summarized by the temperature $T$ and the flow velocity $u^\mu$, while the \emph{equilibrium} electromagnetic emission rate~\cite{Kapusta:2006pm} can be used for $E d^3 \Gamma/d k^3$:
\begin{equation}
	K^0 \frac{d^3 N}{d k^3}=\int_{V_4} d^4 X \left[ K^0 \frac{d^3 \Gamma_{\textrm{eq}}(K_\mu u^\mu, T)}{d k^3} \right] \;.
	\label{eq:macroscopic_equilibrium}
\end{equation}
where $X$ is the space-time position inside the radiating medium and $V_4$ delimits the four-volume of the medium.
The equilibrium rate $E d^3 \Gamma_{\textrm{eq}}/d k^3$ has been computed in many different limits, including some studies
with lattice methods (see Ref.~\cite{BalaProc} and references therein).

\paragraph{Macroscopic approach --- out of equilibrium}

One challenge of the ``macroscopic'' approach is when the medium deviates from local equilibrium, as is the case in viscous hydrodynamics simulations of heavy-ion collisions. A non-negligible effect of viscosity implies that the timescale necessary for the plasma to return to equilibrium ($\tau_{\textrm{relax}}$) is not small compared to the timescale over which the medium changes appreciably ($\tau_{\textrm{evolution}})$: $\tau_{\textrm{relax}} \nll \tau_{\textrm{evolution}}$.
That is, the expanding plasma is constantly pushing the system out of local equilibrium. This out-of-equilibrium evolution leads to a non-negligible increase in the entropy of the system. One expects more photons and dileptons to be produced in this non-equilibrium case than in the local equilibrium case, as a result of the additional entropy generated. Determining the exact energy and momentum distribution of these additional photons and dileptons is not trivial. Although general approaches have been put forward in the past~\cite{Serreau:2003wr,Liu:2017fib}, it is more common to return to a microscopic (quasiparticle) picture of photon and dilepton production: the viscous energy-momentum tensor is related to a certain form for the momentum distribution of quarks and gluons, or of hadrons\footnote{This is the same challenge as the so-called ``viscous corrections'' for particlization into hadrons at the end of hydrodynamic simulations.} (``Grad'', ``Chapman-Enskog'', ``Romatschke-Strickland'', \ldots), and a non-equilibrium electromagnetic emission rate is computed (see Refs~\cite{Hauksson:2017udm,Vujanovic:2019yih,Kasmaei:2019ofu} and references therein). 
The resulting formula used to compute photon and dilepton emission in the non-equilibrium case is effectively the same as Eq.~\ref{eq:macroscopic_equilibrium}, with
\begin{equation}
K^0 \frac{d^3 \Gamma_{\textrm{eq}}(K_\mu u^\mu, T)}{d k^3} \to K^0 \frac{d^3 \Gamma_{\textrm{non-eq}}(K_\mu u^\mu, T,\Pi,\pi^{\mu\nu})}{d k^3}
\label{eq:non_eq_rate}
\end{equation}
where the viscous part of the energy-momentum tensor was decomposed into the shear stress $\pi^{\mu\nu}$ component and the bulk pressure $\Pi$ component. 

\section{Photons and dileptons from hard interactions}

\label{sec:hard}

\paragraph{Binary scaling} At the initial impact of the nuclei, quarks and gluons from each nucleus interact and produce photons (e.g. $q \bar{q} \to g \gamma$) and dileptons  (e.g. $q \bar{q} \to l^+ l^-$). This is the dominant process of non-hadronic-decay production of high-energy photons and high-invariant-mass dileptons. These photons and dileptons provide an important bridge between proton-proton collisions and nucleus-nucleus collisions: their production in nucleus-nucleus collisions is simply an incoherent superposition of nucleon-nucleon collisions, e.g. (for photons):
\begin{equation}
\frac{d N^{\gamma}_{A A}/d p_T^\gamma}{d N^{\gamma}_{p p}/d p_T^\gamma} \approx  N_{\textrm{binary}} \;\;\; \textrm{ for $p_T^\gamma \gtrsim 5$--$10$ GeV}
\label{eq:binary_scaling}
\end{equation}
where $N_{\textrm{binary}}$ is the number of nucleon-nucleon collisions occurring in the nucleus-nucleus collision, which can be calculated with the Glauber model~\cite{Miller:2007ri}. A similar ``binary scaling'' is observed for the production of $Z$ and $W^\pm$ bosons. Equation~\ref{eq:binary_scaling} can be interpreted in the following way~\cite{Arleo:2011gc}:
\begin{align}
\frac{d N^{\gamma}_{p p}}{d p_T^\gamma}=\frac{1}{\sigma^{\textrm{inel}}_{pp}} & f_{a/p} \otimes f_{b/p} \otimes \left[ d\hat{\sigma}_{a b \to c  \gamma} + (\textrm{fragmentation component}) \right]  \label{eq:cross-section_pp} \\
\frac{d N^{\gamma}_{A A}}{d p_T^\gamma}=\frac{N_{\textrm{binary}}}{\sigma^{\textrm{inel}}_{pp}} & f_{a/A} \otimes f_{b/A} \otimes \left[ d\hat{\sigma}_{a b \to c  \gamma} + (\textrm{fragmentation component}) \right]
\label{eq:cross-section_binary_scaling}
\end{align}
where $\otimes$ is a short-hand notation to denote integration over partonic momenta, $f_{a/X}$ are the parton distribution functions for a parton $a$ in a proton/nucleus $X$,  $d \hat{\sigma}$ are partonic cross-sections which can be calculated perturbatively~\cite{Aurenche:2006vj} and $\sigma^{\textrm{inel}}_{pp}$ is the inelastic nucleon-nucleon cross-section. There is a fragmentation component as well, discussed below in more detail, which is subleading at high $p_T^\gamma$ and generally suppressed in measurements with isolation cuts.
The scaling of Eq.~\ref{eq:cross-section_binary_scaling} with the number of binary nucleon-nucleon collisions $N_{\textrm{binary}}$ is assumed to be \emph{exact}. Setting aside fragmentation photons, if Eq.~\ref{eq:binary_scaling} deviates from $1$, it is understood to be due to the different parton content of nuclei compared to protons (a combination of nuclear parton distribution function and the different parton content of protons and neutrons). The production of dileptons (at high enough invariant mass) and $Z$ and $W^\pm$ bosons do not have contributions from fragmentation, and are also used to measure the parton content of nuclei~\cite{Ethier:2020way}.
Equation~\ref{eq:binary_scaling} also been used to \emph{measure} the number of binary collisions $N_{\textrm{binary}}$~\cite{DreesProc}.
Note that the idea that binary scaling is not exact has been explored in the literature (see Ref.~\cite{Eskola:2020lee} and references therein).

\paragraph{Low momentum and invariant mass limit}

A crucial question regarding Equations~\ref{eq:cross-section_pp} and \ref{eq:cross-section_binary_scaling} is: how low in transverse momentum can they be used? When Equation~\ref{eq:cross-section_pp} is compared with data from proton-proton collisions, it is understood that the photon $p_T^\gamma$ must be at least a few GeV for the perturbative Equation~\ref{eq:cross-section_pp} to be valid; photons with $p_T^\gamma \lesssim 1$--$5$~GeV are assumed to be produced non-perturbatively, or at least have significant non-perturbative corrections, though calculations of these non-perturbative processes are generally not available. 

Evidently, any limitation of Equation~\ref{eq:cross-section_pp} in proton-proton collisions applies to Equation~\ref{eq:cross-section_binary_scaling} in heavy-ion collisions. This is a challenge, since the most interesting photon signal in heavy-ion collisions is at low $p_T^\gamma$. The analogous challenge for dileptons is the low-invariant-mass limit of Drell-Yan dileptons. A better  understanding of these low-$p_T^\gamma$/low-invariant-mass regimes in proton-proton collisions would benefit the study of electromagnetic probes in heavy-ion collisions.

\paragraph{Modification of hard scatterings in heavy-ion collisions}

Following the notation of Eqs~\ref{eq:cross-section_pp} and \ref{eq:cross-section_binary_scaling}, the production of high-energy hadrons of species $h$ in proton-proton collisions and heavy-ion collisions is given by
\begin{align}
	\frac{d N^{h}_{p p}}{d p_T^h}=\frac{1}{\sigma^{\textrm{inel}}_{pp}} &f_{a/p} \otimes f_{b/p} \otimes  d\hat{\sigma}_{a b \to c d} \otimes D_{c/h} \label{eq:cross-section_pp_jet} \\
	\frac{d N^{h}_{A A}}{d p_T^h}=\frac{N_{\textrm{binary}}}{\sigma^{\textrm{inel}}_{pp}}  & f_{a/A} \otimes f_{b/A} \otimes  d\hat{\sigma}_{a b \to c d} \otimes (\textrm{parton energy loss})\otimes D_{c/h} 
	\label{eq:cross-section_binary_scaling_jet}
\end{align}
where $D_{c/h}$ is the hadron fragmentation function. We use ``(parton energy loss)'' as a short-hand for the complex energy loss process through which hard partons interact with the lower-energy bath of quarks and gluons. Hard partons lose more or less energy depending on their exact trajectory in the space-time profile (Figure~\ref{fig:temperature_profile}) of the quark-gluon plasma. This combination of the momentum-space formulation of Eq.~\ref{eq:cross-section_pp_jet} with the space-time dependence of parton energy loss is a significant challenge of Eq.~\ref{eq:cross-section_binary_scaling_jet}; in practice, Eq.~\ref{eq:cross-section_binary_scaling_jet} is generally solved with Monte Carlo event generators. For photons (and dileptons), the ``(parton energy loss)'' part of Eq.~\ref{eq:cross-section_binary_scaling_jet} already has one consequence: the processes that lead to parton energy loss will also produce additional photons and dileptons, known as ``jet-medium photons and dileptons''. Recent results are reported in these proceedings~\cite{RouzProc}, and show a non-negligible production of jet-medium photons at intermediate $p_T^\gamma$.

The second consequence of parton energy loss can be seen by rewriting Eqs~\ref{eq:cross-section_pp} and \ref{eq:cross-section_binary_scaling} to include the fragmentation component:\begin{align}
	\frac{d N^{\gamma}_{p p}}{d p_T^\gamma}=\frac{1}{\sigma^{\textrm{inel}}_{pp}} & f_{a/p} \otimes f_{b/p} \otimes \left[ d\hat{\sigma}_{a b \to c  \gamma} + d\hat{\sigma}_{a b \to c d} \otimes D_{c/\gamma} \right]  \label{eq:cross-section_pp_photon_with_frag} \\
	\frac{d N^{\gamma}_{A A}}{d p_T^\gamma}=\frac{N_{\textrm{binary}}}{\sigma^{\textrm{inel}}_{pp}} & f_{a/A} \otimes f_{b/A} \otimes \left[ d\hat{\sigma}_{a b \to c  \gamma} + d\hat{\sigma}_{a b \to c d} \otimes (\textrm{parton energy loss}) \otimes D_{c/\gamma} \right]
	\label{eq:cross-section_binary_scaling_photon_with_frag}
\end{align}
where $D_{c/\gamma}$ is the photon fragmentation function.
Parton energy loss modifies the production of fragmentation photons. Overall, the combination of additional jet-medium radiation and suppressed fragmentation photons remains a remarkably difficult phenomenon to model, combining  all the challenges of parton energy loss with that of low-energy parton and photon production.

\section{Pre-equilibrium photons and dileptons}

\label{sec:pre-eq}

If we think of the applicability of hydrodynamics in terms of the relaxation time of the medium's constituents $\tau_{\textrm{relax}}$ and the evolution time of the medium $\tau_{\textrm{evolution}}$, we can broadly categorize the early stage of the collisions in the three regimes shown in Table~\ref{tab:stages} (see Ref.~\cite{BoguslavskiProc} for a broader discussion of the early stage of heavy-ion collisions and for references).
\begin{table}[h!]
\centering
\begin{tabular}{|c|c|c|}
	\hline
	Earliest times & Before hydro. initialization time & After hydro. initialization time \\
	\hline
	$\tau_{\textrm{relax}} \gg \tau_{\textrm{evolution}}$	& $\tau_{\textrm{relax}} \sim \tau_{\textrm{evolution}}$  & $\tau_{\textrm{relax}} \ll \tau_{\textrm{evolution}}$  \\
	\hline
	Expansion dominated & Transition & Interaction dominated \\
	\hline
	``Free-streaming regime'' & ``Hydrodynamization'' & ``Hydrodynamics regime'' \\
	\hline
\end{tabular}
\caption{Schematic classification of the different regimes of the early stage of heavy-ion collisions.}
\label{tab:stages}
\end{table}

Photons and dileptons can be produced in all stages, and a consistent description of these different stages is necessary to calculate accurately electromagnetic emission in heavy-ion collisions. 
One approach to study pre-equilibrium photons and dileptons is to focus on conformal boost-invariant systems where inhomogeneities in the plane transverse are neglected; this leads to simplified $0+1$D systems where photon and dilepton production can be calculated in great detail using kinetic theory (see Refs~\cite{CoquetProc,PlaschkeProc} and references therein) and holography~\cite{Grieninger:2022yps}.
In general, computing pre-equilibrium photons and dileptons in heavy-ion collisions require additional modeling beyond conformal $0+1$D systems, which have been explored in the past~\cite{Greif:2016jeb, Oliva:2017pri, Berges:2017eom,Monnai:2019vup,Churchill:2020uvk,Garcia-Montero:2019vju,Bhattacharya:2015ada, Gale:2021emg}.

\section{Hydrodynamic regime}

\label{sec:hydro}

\paragraph{Thermometer and chromometer}
Electromagnetic radiation produced in the hydrodynamic regime of the plasma are known as \emph{thermal} photons and dileptons. We expect them to provide direct information on the properties of the quark-gluon plasma, given that their rate of emission and their momentum distribution reflect the local properties of the degrees of freedom that radiated them. Thermal photons and dileptons are sometimes referred to as ``chronometers'' and ``thermometers'' of the plasma, although they face stiff competition from hadrons in that regard.
Hydrodynamic simulations of heavy-ion collisions, which are overwhelmingly calibrated on \emph{hadronic} measurements, provide strong constraints on the temperature profile of the plasma; Figure~\ref{fig:temperature_profile} is an example, entirely calibrated using hadronic data. Similarly, the lifetime of the quark-gluon plasma, inasmuch as such a number can be defined in a system without a phase transition, is already known from hydrodynamic simulations of heavy-ion collisions (it can be read off directly from Figure~\ref{fig:temperature_profile}, for example, given a certain definition of ``lifetime'').

On the other hand, by focusing on certain bins of momentum and invariant mass and by subtracting certain hadronic decays, it is possible for photons and dileptons to provide information on the earlier stage of the collisions, to which hadrons have limited sensitivity. Moreover, photons and dileptons produced in hadronic interactions provide a valuable window into the finite-density properties of these hadrons (see Section~\ref{sec:hadronic}).

\paragraph{Photons and dileptons as probes of the quark-gluon plasma}

Thermal photons and dileptons are calculated according to Eqs~\ref{eq:macroscopic_equilibrium} and \ref{eq:non_eq_rate}. Because hadrons are produced at low temperature, their strongest constraints on the temperature profile of heavy-ion collisions are at these low temperatures (around the $T=150$~MeV contour of Figure~\ref{fig:temperature_profile}). Photons and dileptons are more sensitive to the higher temperature regions of the plasma, and can help better constrain the space-time profile of the plasma and its  properties: as more measurements, and in particular more precise ones, become available, electromagnetic probes will provide increasingly strong constraints on the chemical and thermal equilibration of the plasma (see Section~\ref{sec:pre-eq}), the temperature dependence of the shear and bulk viscosities, potentially of higher-order transport coefficients~\cite{Vujanovic:2016anq}, and more. There have also been proposals to use the polarization of photons and dileptons to increase measurement sensitivity to the system's momentum anisotropy; see Ref.~\cite{HaukssonProc} and references therein.


\section{Photons from hadronic interactions and hadronic decays}

\label{sec:hadronic}

\paragraph{Hadronic interactions and decays at finite temperature/density}

The quark-gluon plasma can reconfine into hadrons through a crossover (at low baryon chemical potential) or through a possible phase transition (at higher baryon chemical potential). In either case, for some time after confinement, the density of hadrons will remain high, and their properties will be different from their vacuum properties.
More pronounced ``finite-density'' effects are seen on short-lived mesons, such as the $\rho$ and $a_1(1260)$ chiral partners, whose evolution during the crossover transition to deconfined quarks and gluons informs us about how chiral symmetry is restored by deconfinement~\cite{Geurts:2022xmk}; this can be studied at lower temperature and higher densities as well~\cite{TripoltProc}.
Because short-lived hadrons contribute to both photon and dilepton production through hadronic interactions and decays, they provide invaluable information on the finite-temperature properties of hadrons and the transition to deconfined quarks and gluons. Readers are referred to Ref.~\cite{Geurts:2022xmk} for a recent comprehensive review.

\paragraph{Photons and dileptons from microscopic description of hadronic interactions}

The transition from hydrodynamics to hadronic transport in simulations of heavy-ion collisions can also be understood in terms of the relaxation time $\tau_{\textrm{relax}}$ and the evolution time $\tau_{\textrm{evolution}}$ discussed earlier. When quarks and gluons reconfine into hadrons, only a residual nuclear interaction strength remains, much smaller than the interaction between deconfined quarks and gluons, leading to a larger relaxation time $\tau_{\textrm{relax}}$. Thus, in the later stage of heavy-ion collisions, while the evolution timescale $\tau_{\textrm{evolution}}$ is not changing very rapidly, the relaxation time $\tau_{\textrm{relax}}$ increases suddenly. This effectively leads to the inverse processes listed in Table~\ref{tab:stages}: hydrodynamics transitioning towards free-streaming.

Computing photons and dileptons production in the transition from hydrodynamics to free-streaming hadrons is challenging. It is still common to use a \emph{macroscopic} approach (Eqs~\ref{eq:macroscopic_equilibrium} and \ref{eq:non_eq_rate}) to perform this calculation: the hydrodynamic simulation is continued to later times, and combined with a photon or a dilepton emission rate. However, important work has been done to compute electromagnetic emission microscopically instead, using hadronic transport (see Ref.~\cite{ElfnerProc} and references therein). This provides the great benefit of unifying the description of electromagnetic probes with that of hadrons, which have been using hadronic transport at the end of hydrodynamic simulations for years.

\section{Photons and dileptons at low collision energies}

The discussion of all the different sources of photons and dileptons presented above is not only relevant for high-energy nuclear collisions (LHC and top RHIC collision energies), but also for lower energy collisions studied in the beam energy scan at RHIC, as well as the HADES and future FAIR experiment at GSI. Many of the challenges discussed above are amplified in lower collision energies. In particular, the quark-gluon plasma phase is shorter relative to the complex early stage of the collisions. Calculations of the contribution of hard processes (Section~\ref{sec:hard}) is also more difficult due to the lower center-of-mass energies. Calculations of photons using a hydrodynamic-based model are reported in Ref.~\cite{ShenProc}; they show the potential of photons to help understand the early stage of the collisions. Electromagnetic production in even lower collision energies is being studied as well, see e.g. Refs~\cite{Savchuk:2022aev,Seck:2020qbx}.

\section{Outlook}

Because photons and dileptons can be produced at any stage of heavy-ion collisions, progress made in understanding almost every aspect of heavy-ion collisions has direct connections with electromagnetic probes. The early stage of heavy-ion collisions remains challenging to describe, and increasingly accurate measurements of electromagnetic probes will provide valuable information to help understand it. More generally, photons and dileptons can help better understand the transition to and from the hydrodynamics regime, which is particularly important at lower collision energies. Photons and dileptons further provide a valuable tool to study the temperature and baryon chemical potential dependence of the equation of state and transport coefficients.

\paragraph{Acknowledgments} 

I thank Gojko Vujanovic for valuable feedback, and the participants in the KITP workshop ``The Many Faces of Relativistic Fluid Dynamics'' for many discussions that guided part of these proceedings. This research was supported in part by the National Science Foundation under Grant No. NSF PHY-1748958.

\bibliographystyle{JHEP}
\bibliography{inspire,noninspire}

\end{document}